\documentclass[apjl]{emulateapj}

\usepackage{txfonts}
\usepackage{graphicx}
\usepackage{amssymb}
\usepackage[below]{placeins}
\usepackage{txfonts}
\usepackage{natbib}
\bibpunct{(}{)}{;}{a}{}{,}
\def \arcmin{$^{\prime}$}

\def \chandra{{\emph{Chandra}}}

\bibpunct{(}{)}{;}{a}{}{,}

\shorttitle{Filamentary cold gas in M~87}
\shortauthors{Werner et al.}

\begin{document}

\title{The nature of filamentary cold gas in the core of the Virgo Cluster}

\author{N. Werner\altaffilmark{1}, J. B. R. Oonk\altaffilmark{2}, R.~E.~A.~Canning\altaffilmark{1}, S.~W.~Allen\altaffilmark{1,3}, A. Simionescu\altaffilmark{1}, J.~Kos$\altaffilmark{2,4}$, R.~J.~van~Weeren\altaffilmark{5}, A.~C.~Edge\altaffilmark{6},  
A.~C.~Fabian\altaffilmark{7}, A.~von~der~Linden\altaffilmark{1}, P.~E.~J.~Nulsen\altaffilmark{5}, C.~S.~Reynolds\altaffilmark{8}, M.~Ruszkowski\altaffilmark{9,10} }
\affil{$^1$Kavli Institute for Particle Astrophysics and Cosmology, Stanford University, 452 Lomita Mall, Stanford, CA 94305-4085, USA; \\ e-mail: norbertw@stanford.edu}
\affil{$^2$ASTRON, Netherlands Institute for Radio Astronomy, P.O. Box 2, 7990 AA Dwingeloo, The Netherlands}
\affil{$^3$SLAC National Accelerator Laboratory, 2575 Sand Hill Road, Menlo Park, CA 94025, USA}
\affil{$^4$Faculty of Mathematics and Physics, University of Ljubljana, Jadranska 19, 1000 Ljubljana, Slovenia}
\affil{$^5$Harvard-Smithsonian Center for Astrophysics, 60 Garden Street, Cambridge, MA 02138, USA}
\affil{$^6$Institute for Computational Cosmology, Department of Physics, Durham University, Durham, DH1 3LE, UK}
\affil{$^7$Institute of Astronomy, Madingley Road, Cambridge CB3 0HA, UK}
\affil{$^8$Department of Astronomy and the Maryland Astronomy Center for Theory and Computation, University of Maryland, College Park, MD 20742, USA}
\affil{$^9$Department of Astronomy, University of Michigan, 500 Church Street, Ann Arbor, MI 48109, USA}
\affil{$^{10}$Michigan Center for Theoretical Physics, 3444 Randall Lab, 450 Church St, Ann Arbor, MI 48109, USA}

\begin{abstract}
We present a multi-wavelength study of the emission-line nebulae located $\sim$38 arcsec (3~kpc in projection) southeast of the nucleus of M~87, the central dominant galaxy of the Virgo Cluster. We report the detection of far-infrared (FIR) 
[\ion{C}{2}] line emission at 158~$\mu$m from the nebulae using observations made with the {\it Herschel} Photodetector Array Camera \& Spectrometer (PACS). The infrared line emission is extended and cospatial with optical H$\alpha$+
[\ion{N}{2}], far-ultraviolet \ion{C}{4} lines, and soft X-ray emission. The filamentary nebulae evidently contain multi-phase material spanning a temperature range of at least 5 orders of magnitude, from $\sim100$~K to $\sim10^7$~K. This 
material has most likely been uplifted by the Active Galactic Nucleus from the center of M~87. The thermal pressure of the $10^4$~K phase appears to be significantly lower than that of the surrounding hot intra-cluster medium (ICM) indicating the presence of additional turbulent and magnetic pressure in the filaments. If the turbulence in the filaments is subsonic then the magnetic field strength required to balance the pressure of the surrounding ICM is $B\sim30-70~\mu$G. The spectral properties of the soft X-ray emission from the filaments indicate that it is due to thermal plasma with $kT\sim0.5$--1~keV, which is cooling by mixing with the cold gas and/or radiatively. Charge exchange can be ruled out as a significant source of soft X-rays. Both cooling and mixing scenarios predict gas with a range of temperatures. This is at first glance inconsistent with the apparent lack of X-ray emitting gas with $kT<0.5$~keV. However, we show that the missing very soft X-ray emission could be absorbed by the cold gas in the filaments with an integrated hydrogen column density of $N_{\rm H} \sim 1.6\times10^{21}$~cm$^{-2}$, providing a natural explanation for the apparent temperature floor to the X-ray emission at $kT\sim0.5$~keV. The FIR through ultra-violet line emission is most likely primarily powered by the ICM particles penetrating the cold gas following a shearing induced mixing process. An additional source of energy may, in principle, be provided by X-ray photoionization from cooling X-ray emitting plasma. The relatively small line ratio of [\ion{O}{1}]/[\ion{C}{2}]$<7.2$ indicates a large optical depth in the FIR lines. The large optical depth in the FIR lines and the intrinsic absorption inferred from the X-ray and optical data imply significant reservoirs of cold atomic and molecular gas distributed in filaments with small volume filling fraction, but large area covering factor.

\end{abstract}


\section{Introduction}

\begin{table*}
\begin{center}
\caption{Summary of observations and of the properties of the FIR lines integrated over the full 5$\times$5 spaxel (47\arcsec$\times$47\arcsec) field of view of {\it Herschel} PACS. }
\begin{tabular}{ccccccc}
\hline\hline
Line & Rest frame $\lambda$ & Observation duration & Line Flux & Instrumental FWHM  & Observed FWHM  & Line shift  \\
         & ($\mu$m) &  (s) &  ($10^{-14}$~erg s$^{-1}$ cm$^{-2}$) & (km~s$^{-1}$) & (km~s$^{-1}$) & (km~s$^{-1}$) \\
\hline
\ion{C}{2} & 157.7 & 3000 & $5.13\pm0.5$ & 240 & $361\pm27$ & $-62\pm11$ \\
\ion{O}{1} &  63.2 & 3312 & $<36.7$\footnote{2$\sigma$ upper limits} & 85  & \ldots & \ldots \\
\ion{O}{1b} & 145.5 & 4480 & $<2.9^{a}$ & 255 & \ldots & \ldots \\
\hline
\vspace{-0.7cm}
\label{obs}
\end{tabular}
\end{center}
\end{table*}

Massive giant elliptical galaxies in the centers of clusters with central cooling times shorter than the Hubble time frequently display spectacular optical emission-line nebulae \citep[e.g.][]
{johnstone1987,heckman1989,donahue1992,crawford1999,mcdonald2010}. 
Associated cold molecular gas has also been detected in many of these systems, from near-infrared (NIR) K-band H$_{2}$ line emission \citep[e.g.][]{jaffe1997,falcke1998,donahue2000,hatch2005,jaffe2005,johnstone2007,oonk2010} and CO observations \citep[e.g.][]{edge2001,edge2003,salome2003,mcdonald2012}. While the NIR spectra show H$_{2}$ line ratios characteristic of collisionally excited 1000--2000~K molecular gas, the CO emission traces the coldest ($<$50~K) molecular gas phases. Optical H$\alpha$+[\ion{N}{2}] line emission arises from a thin ionized skin of 10$^4$~K gas on underlying reservoirs of cold neutral and molecular gas. CO observations indicate that the amounts of molecular gas in cluster cores can be large \citep{edge2001}, e.g. the cold gas mass in the center of the Perseus Cluster is approaching $10^{11}~M_{\odot}$ \citep{salome2006,salome2008b,salome2011}.  
The observed level of star formation in these systems is, however, typically relatively small \citep[e.g.][]{odea2008}. Despite significant efforts to understand the origin and the excitation mechanism of these nebulae, their nature, energy 
source, and detailed physics remain poorly understood.

The relatively nearby \citep[$d\sim$16.7~Mpc,][]{blakeslee2009} giant elliptical galaxy M~87 (NGC~4486), at the centre of the Virgo Cluster, harbors a well-known extended optical H$\alpha$+[\ion{N}{2}] filament system 
\citep{ford1979,sparks1993}. The proximity of this system allows us to study the physics of the emission-line nebulae in greater detail than is possible elsewhere. The H$\alpha$+[\ion{N}{2}] nebulae in M~87 were found to spatially coincide 
with filamentary soft X-ray emission \citep{young2002,sparks2004}, which can be modeled as $\sim$0.5~keV plasma in collisional ionization equilibrium \citep{werner2010}. The emission-line nebulae also spatially coincide with \ion{C}{4} 
line emission at far-ultraviolet wavelengths (FUV), which typically arises in gas at temperature $\sim$10$^5$ K \citep{sparks2009,sparks2012}.  \ion{C}{4} line emission has so far not been detected in extended filaments in any other central 
cluster galaxy.  \citet{werner2010} showed that all of the bright H$\alpha$ and UV filaments in M~87 are found in the downstream region of a $<$3 Myr old shock front revealed by X-ray observations with {\it Chandra} \citep{million2010b}. This 
argues that the generation of H$\alpha$, UV, and soft X-ray emission in M~87 is coupled to shocks in the hot X-ray emitting plasma. Based on these observations \citet{werner2010} proposed that shocks induce shearing around the cooler, 
denser gas filaments, which promotes mixing with the ambient hot intra-cluster medium (ICM) via instabilities. By enhancing the rate at which hot thermal particles come into contact with the colder gas phases, mixing can in principle supply the power and the 
ionizing particles needed to explain the NIR to FUV line emission \citep{ferland2008,ferland2009,fabian2011}. 

The launch of the {\it Herschel} space observatory, with its unprecedented sensitivity to far-infrared (FIR) line emission \citep{pilbratt2010}, has opened new opportunities to study the coldest gas phases in galaxies. {\it Herschel} has enabled 
the first detections of atomic cooling lines of [\ion{C}{2}], [\ion{O}{1}], and [\ion{N}{2}] in the X-ray bright cores of Abell~1068, Abell~2597 \citep{edge2010}, and the Centaurus and Perseus clusters \citep{mittal2011,mittal2012}. These lines are 
the dominant cooling lines for neutral interstellar gas and can be used as diagnostics to infer temperatures, densities, and radiation fields \citep[e.g.][]{kaufman1999}. Because carbon is the fourth most abundant element in the Universe and 
has a low ionization potential, the 158~$\mu$m [\ion{C}{2}] line is the most ubiquitous and best-studied \citep[e.g.][]{malhotra1997,malhotra2001}. It is a tracer of gas with a temperature of $\sim100$~K. 

To study the cold gas phases associated with the filamentary line emission nebulae in M~87 we observed the regions of brightest H$\alpha$+[\ion{N}{2}]  filaments, extending to the southeast of the nucleus of the galaxy, with the {\it Herschel} 
Photodetector Array Camera \& Spectrometer (PACS) at the wavelengths of [\ion{C}{2}]$\lambda157\mu$m, [\ion{O}{1}]$\lambda63\mu$m and [\ion{O}{1b}]$\lambda145\mu$m. 
Here we report the results of these observations, as well as a reanalysis of deep {\it{Chandra}} X-ray, {\it Hubble Space Telescope (HST)} optical and UV data, and long slit spectra obtained with the William Herschel Telescope (WHT). Sect. 
\ref{analysis} describes the observations, data reduction, and analysis of the {\it Herschel} PACS, {\it HST}, {\it Chandra}, and WHT data. In Sect. \ref{results}, we summarize the results, and in Sect.~\ref{discussion} discuss the implications of 
our observations for the nature and the energy sources of the filaments. Our main conclusions are summarized in Sect.~\ref{conclusions}. We assume a distance to M~87 of 16.7~Mpc \citep{blakeslee2009}, which implies a linear scale of 81 
pc arcsec$^{-1}$. The redshift of M~87 is $z= 0.004360$.

\begin{figure}
\begin{minipage}{0.49\textwidth}
\hspace{-1.3cm}\includegraphics[width=11cm,clip=t,angle=180.]{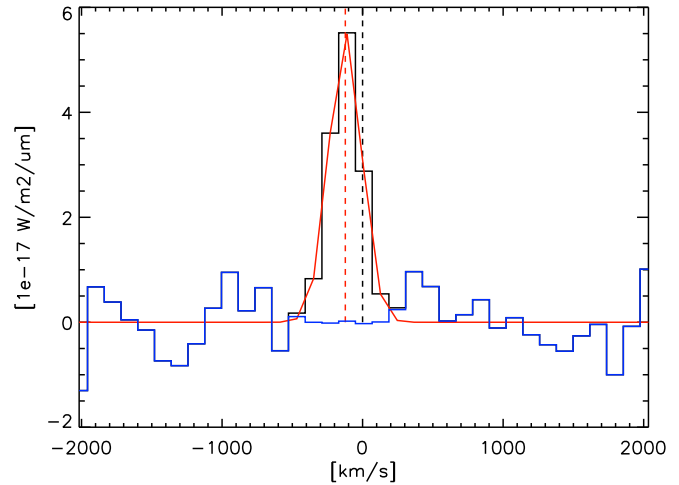}
\vspace{-1cm}
\end{minipage}
\caption{The FIR [\ion{C}{2}]$\lambda157~\mu$m line obtained from the central spaxel ($9.4\times9.4$~arcsec$^2$) of the {\it Herschel} PACS rebinned data cube. The mean line centroid is blue-shifted with respect to M~87 (at 
$z=0.004360$) by $v=-123\pm5$~km~s$^{-1}$. The blue line indicates the detector noise. } 
\label{line}
\end{figure}

\begin{figure*}
\begin{minipage}{0.33\textwidth}
\includegraphics[width=6cm,clip=t,angle=0.]{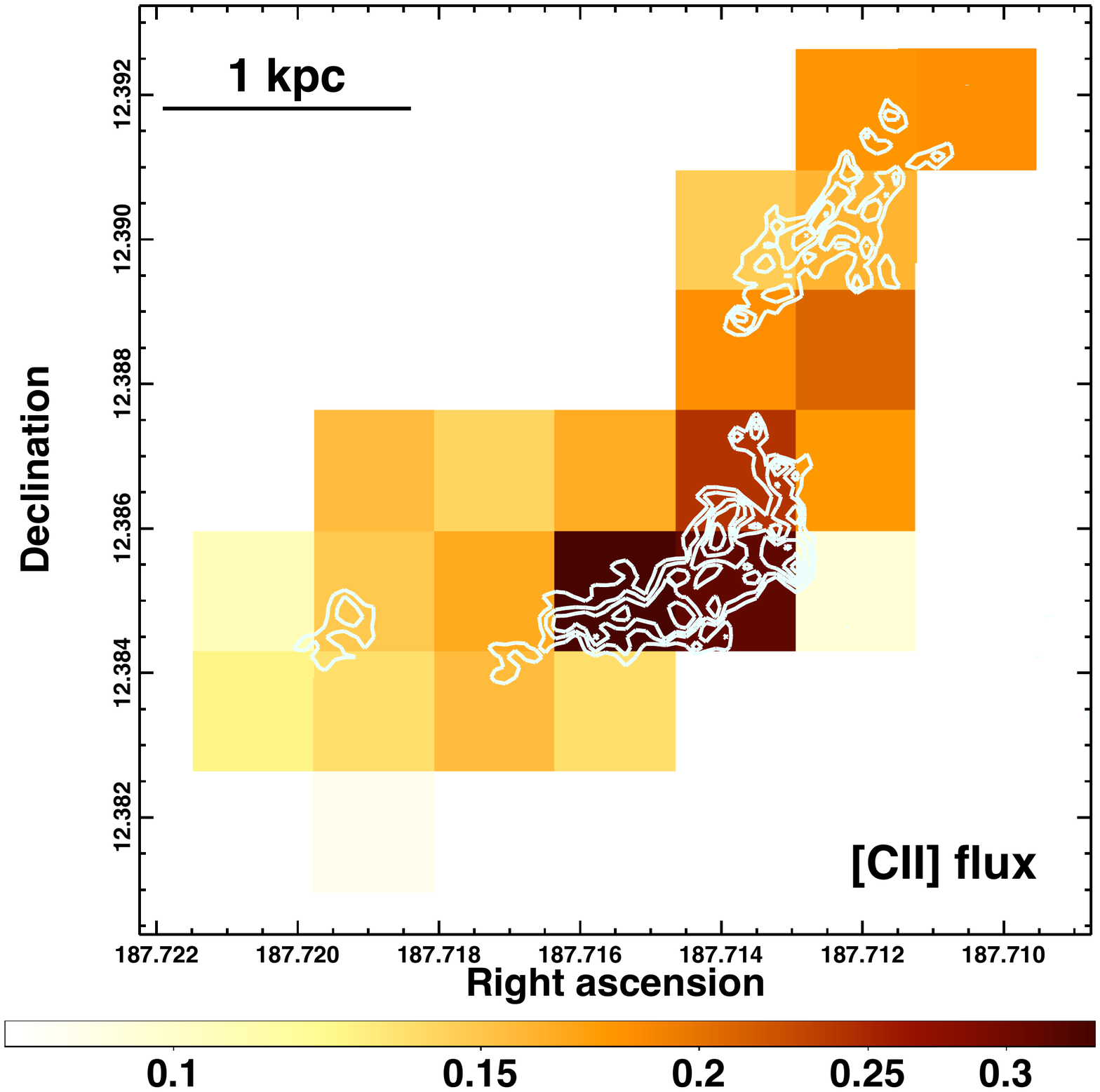}
\end{minipage}
\begin{minipage}{0.33\textwidth}
\hspace{-0.45cm}\hspace{1cm}\includegraphics[width=6cm,clip=t,angle=0.]{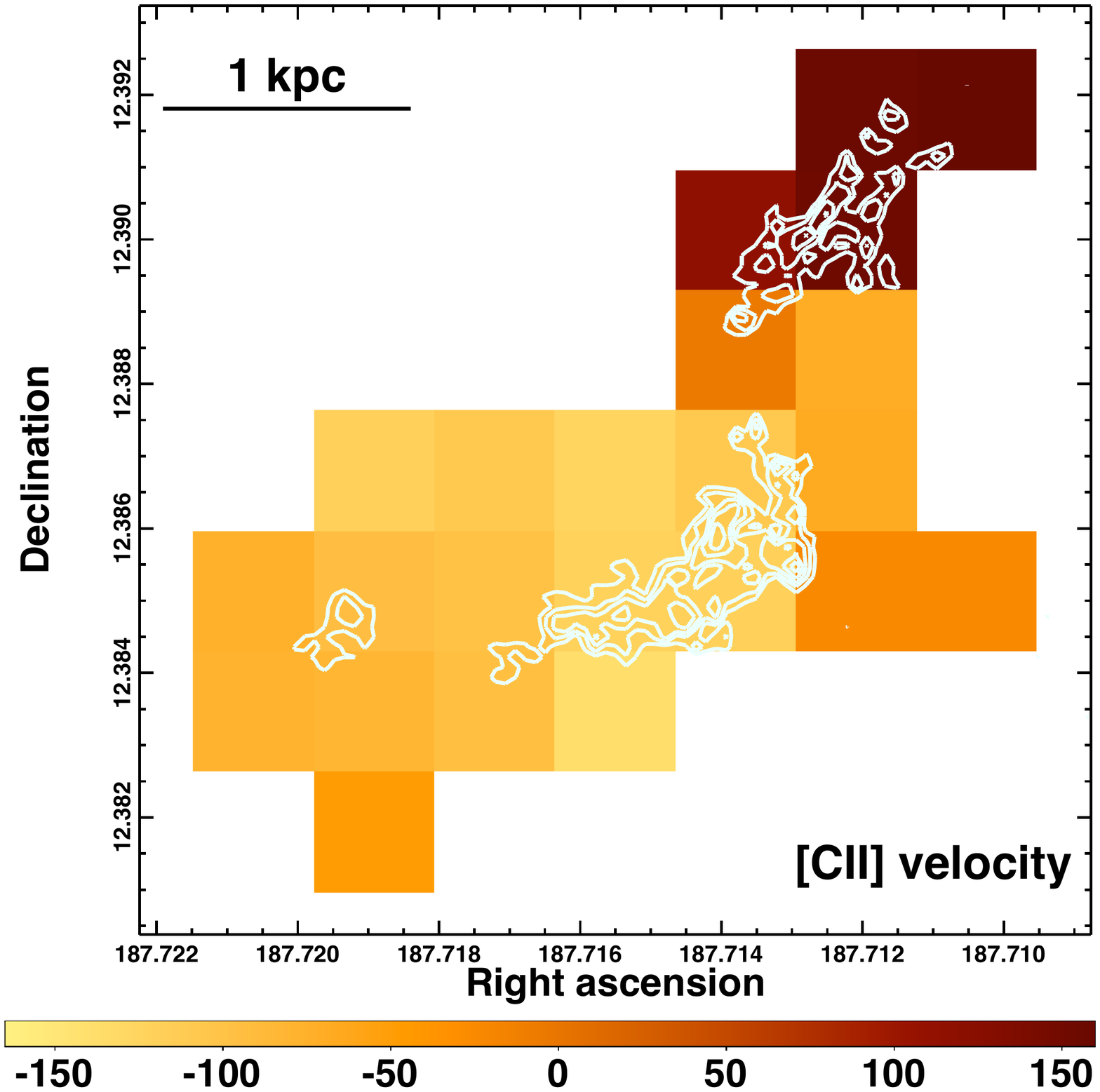}
\end{minipage}
\begin{minipage}{0.33\textwidth}
\hspace{1cm}\includegraphics[width=6cm,clip=t,angle=0.]{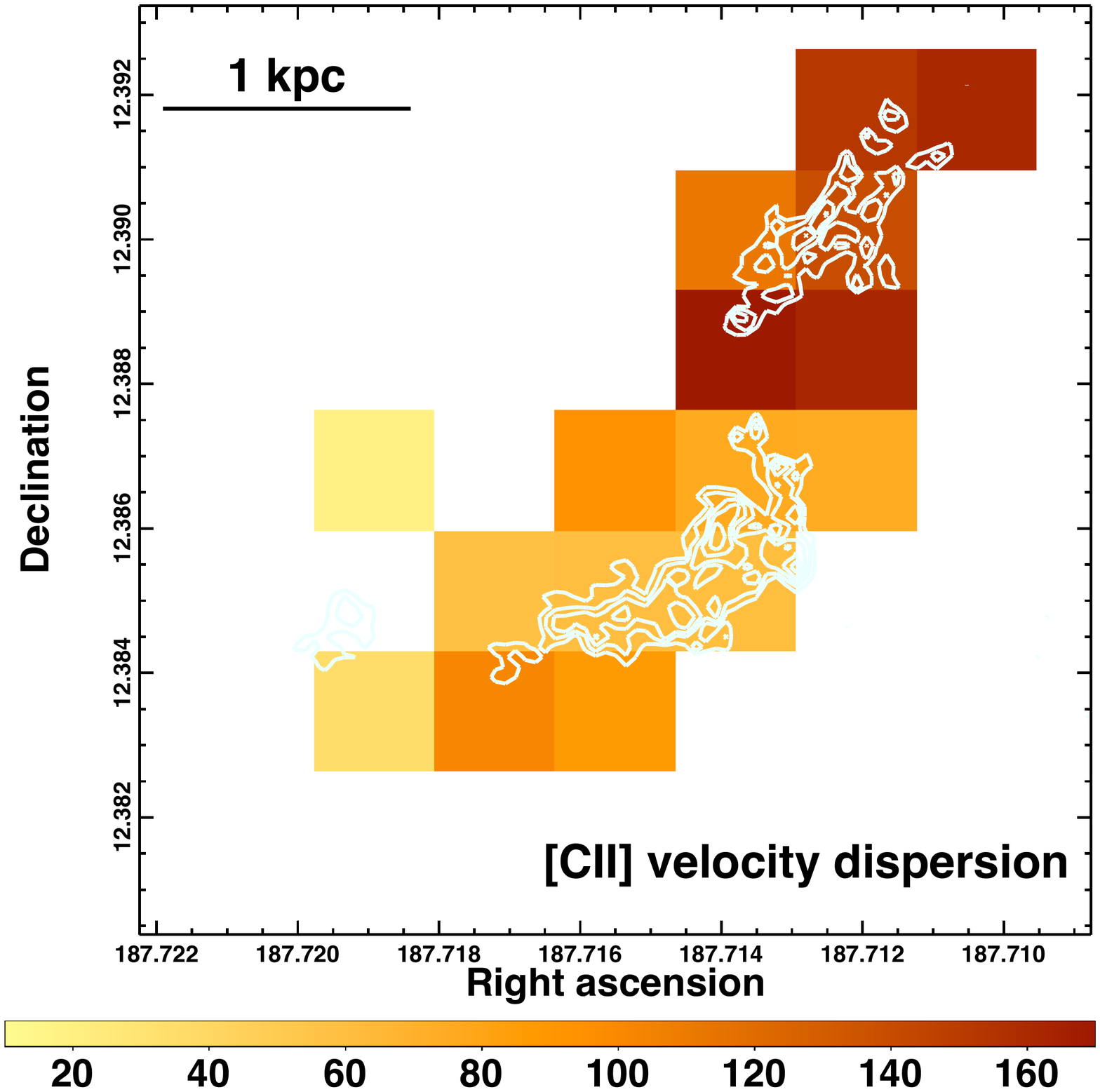}
\end{minipage}
\begin{minipage}{0.33\textwidth}
\vspace{-2cm}
\includegraphics[width=6cm,clip=t,angle=0.]{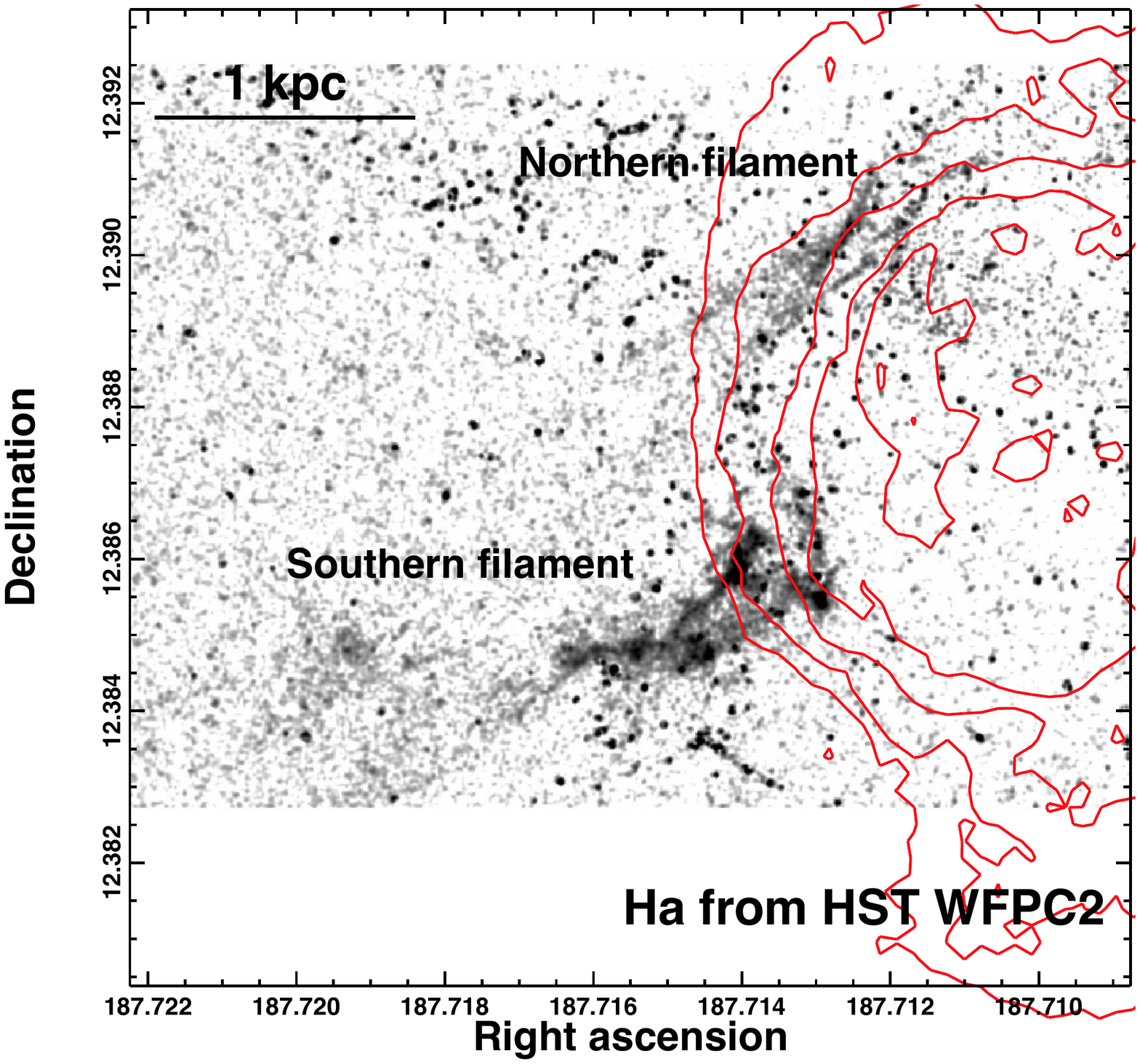}
\end{minipage}
\begin{minipage}{0.33\textwidth}
\vspace{-2cm}
\hspace{-0.45cm}\hspace{1cm}\includegraphics[width=6cm,clip=t,angle=0.]{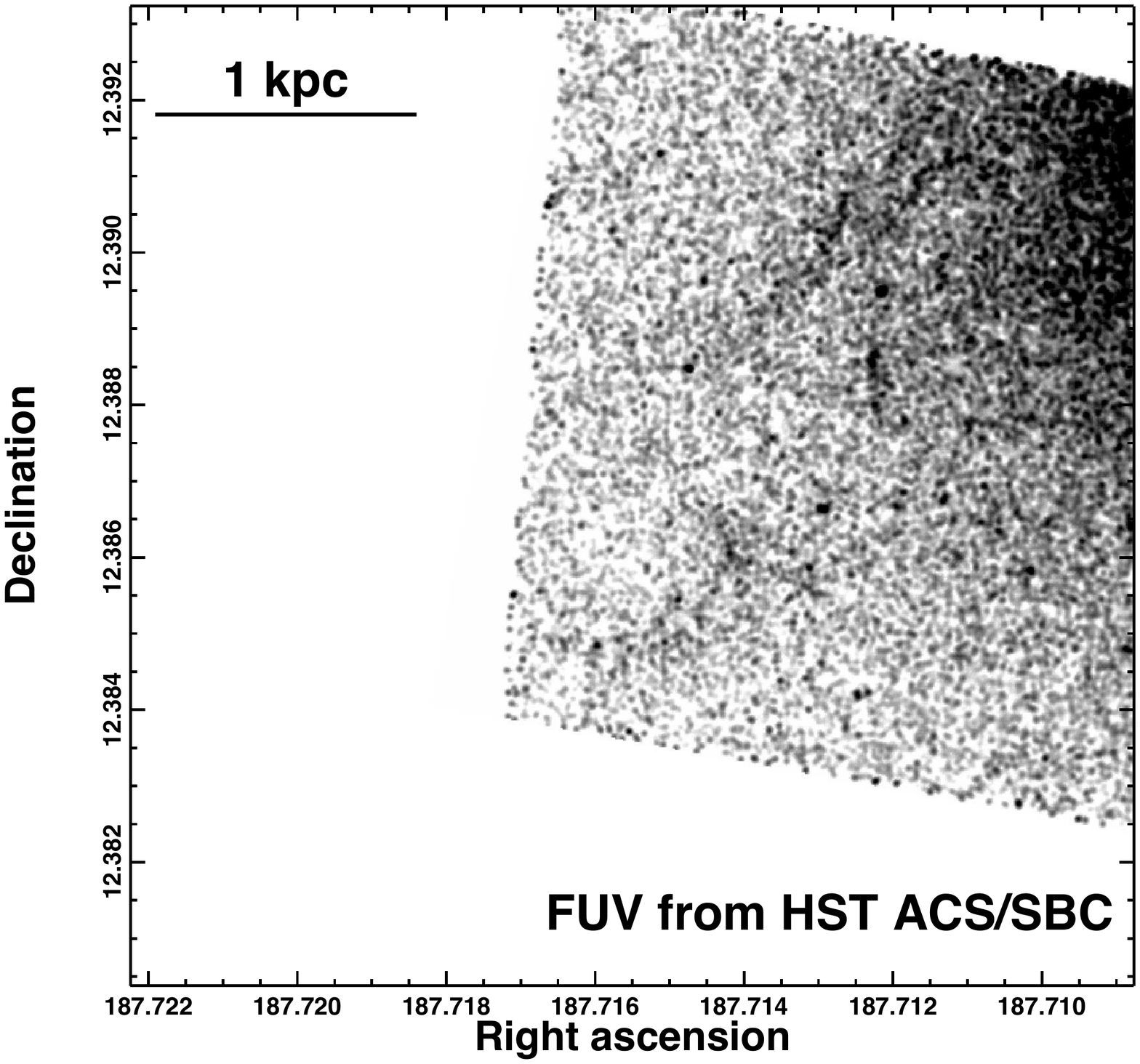}
\end{minipage}
\begin{minipage}{0.33\textwidth}
\vspace{-2cm}
\hspace{1cm}\includegraphics[width=6.2cm,clip=t,angle=0.]{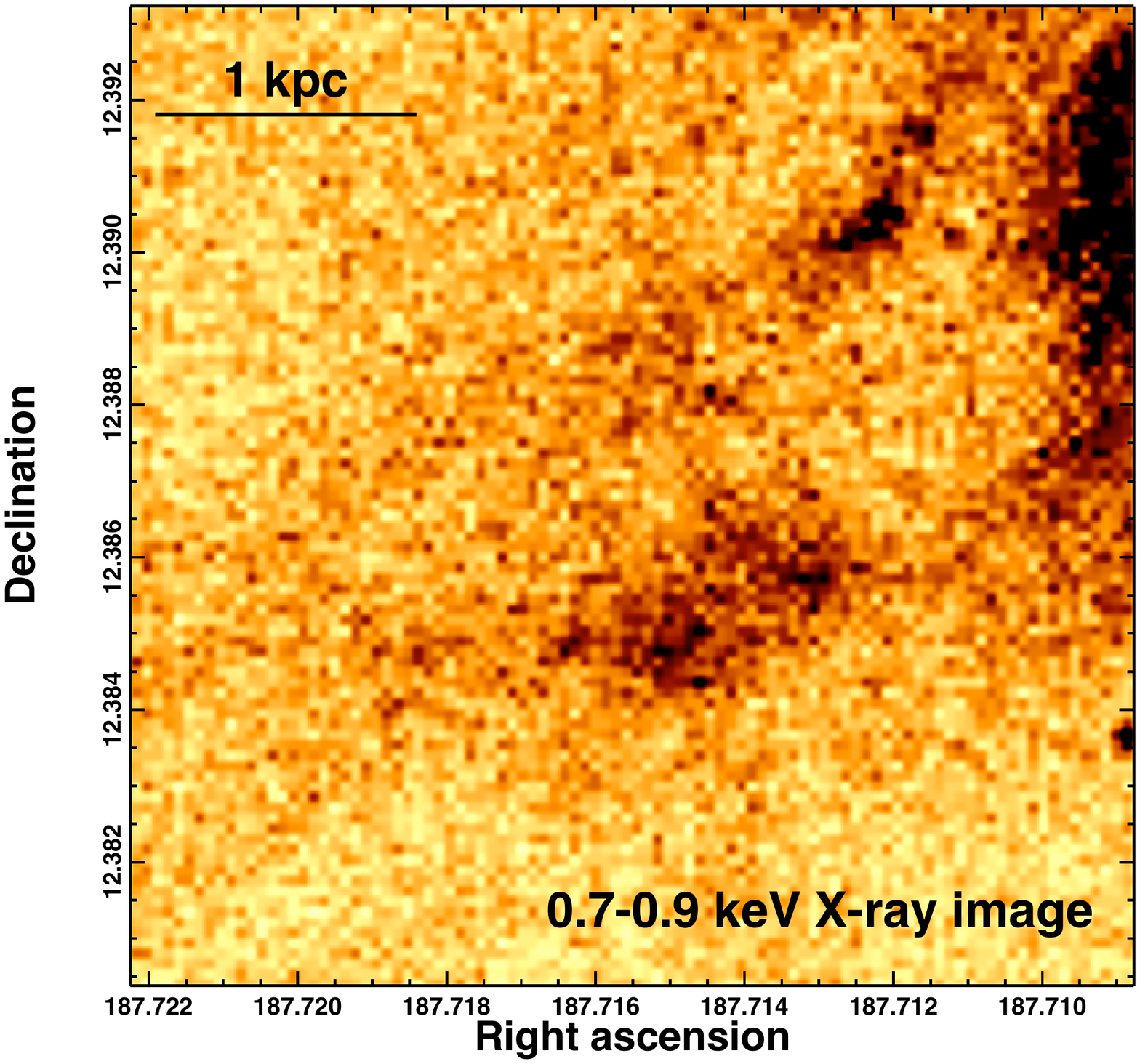}
\end{minipage}
\begin{minipage}{0.33\textwidth}
\vspace{-2cm}
\includegraphics[width=6cm,clip=t,angle=0.]{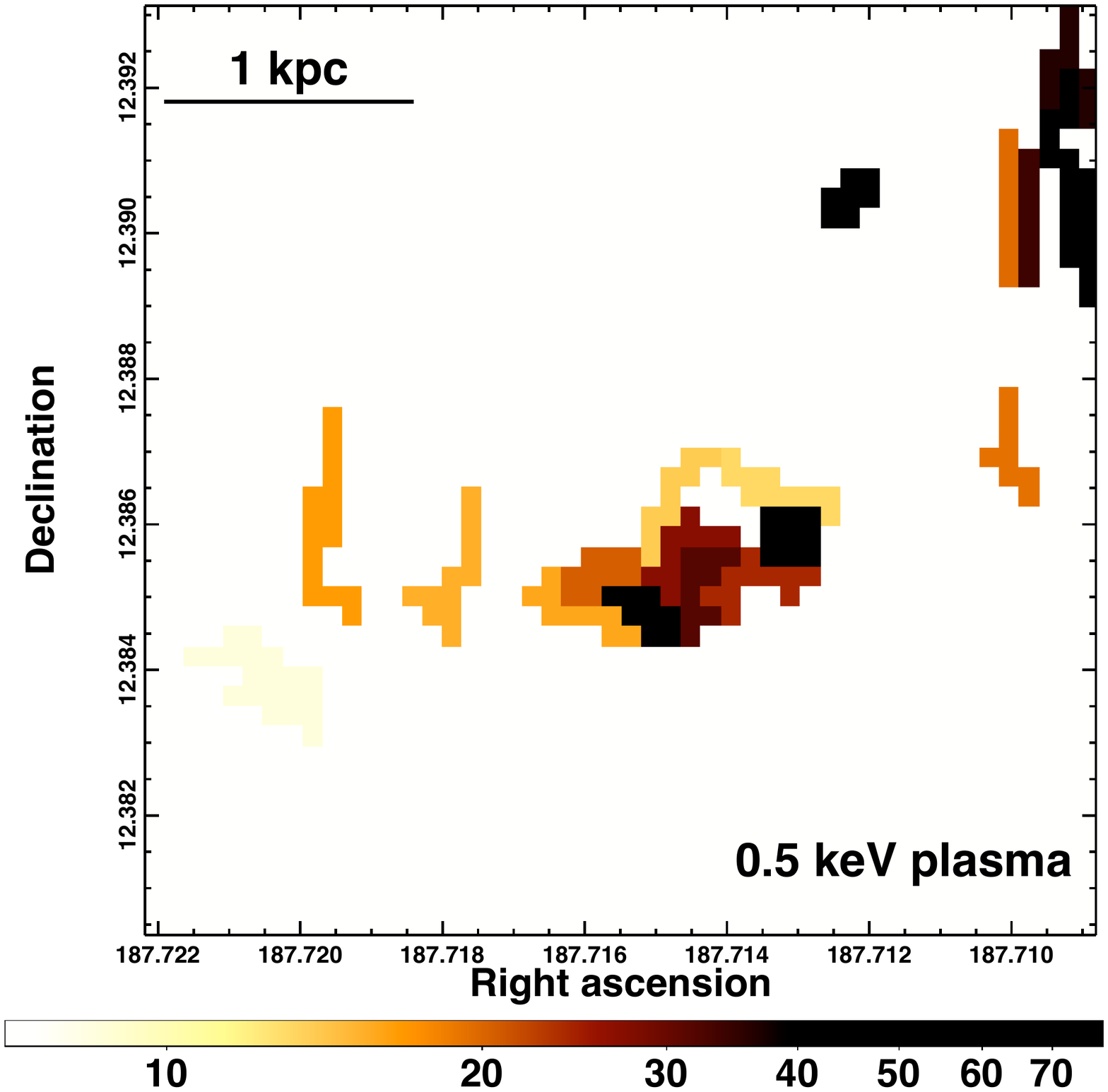}
\end{minipage}
\begin{minipage}{0.33\textwidth}
\vspace{-2cm}
\hspace{-0.45cm}\hspace{1cm}\includegraphics[width=6cm,clip=t,angle=0.]{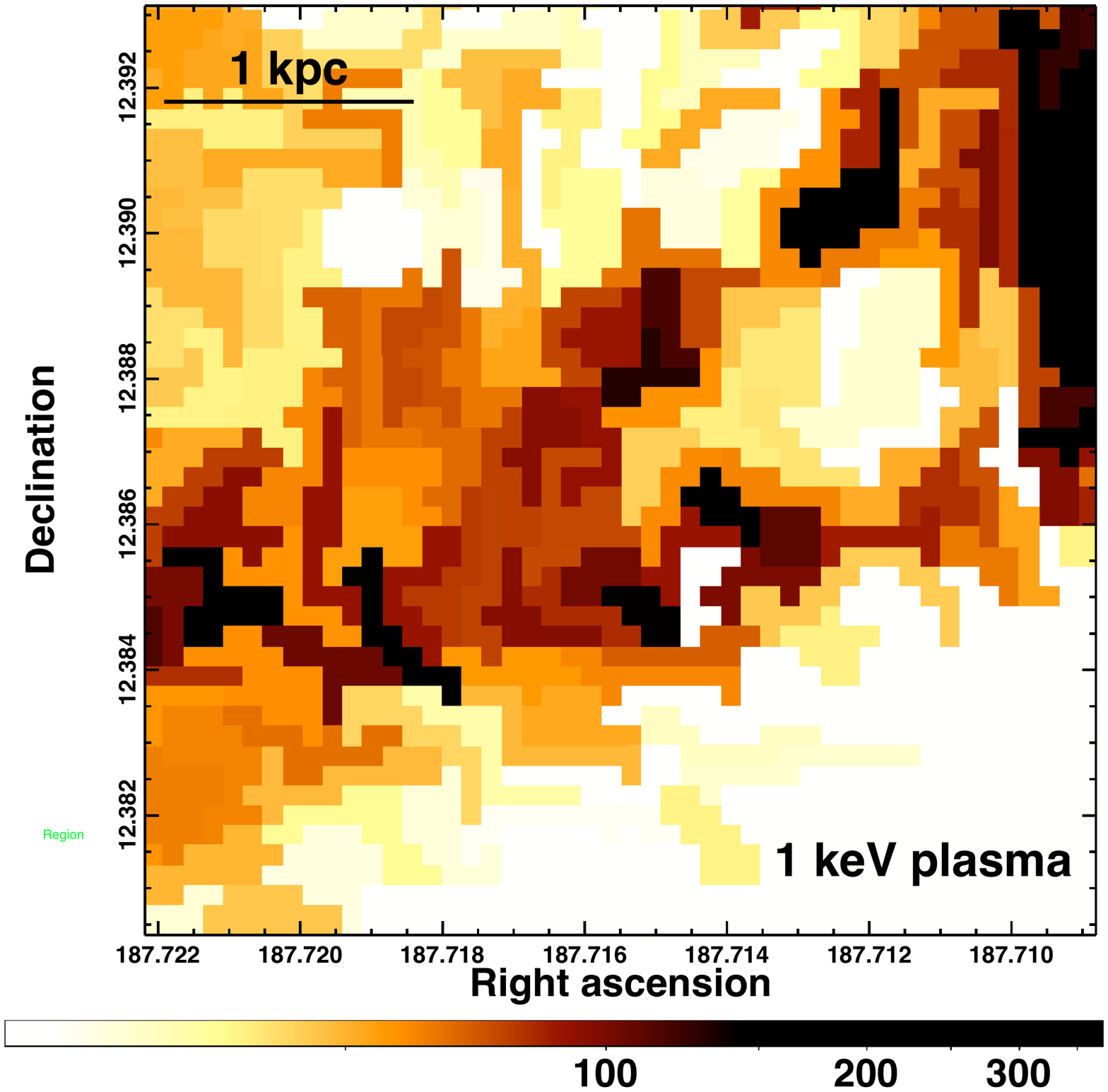}
\end{minipage}
\begin{minipage}{0.33\textwidth}
\vspace{-2cm}
\hspace{1cm}\includegraphics[width=6cm,clip=t,angle=0.]{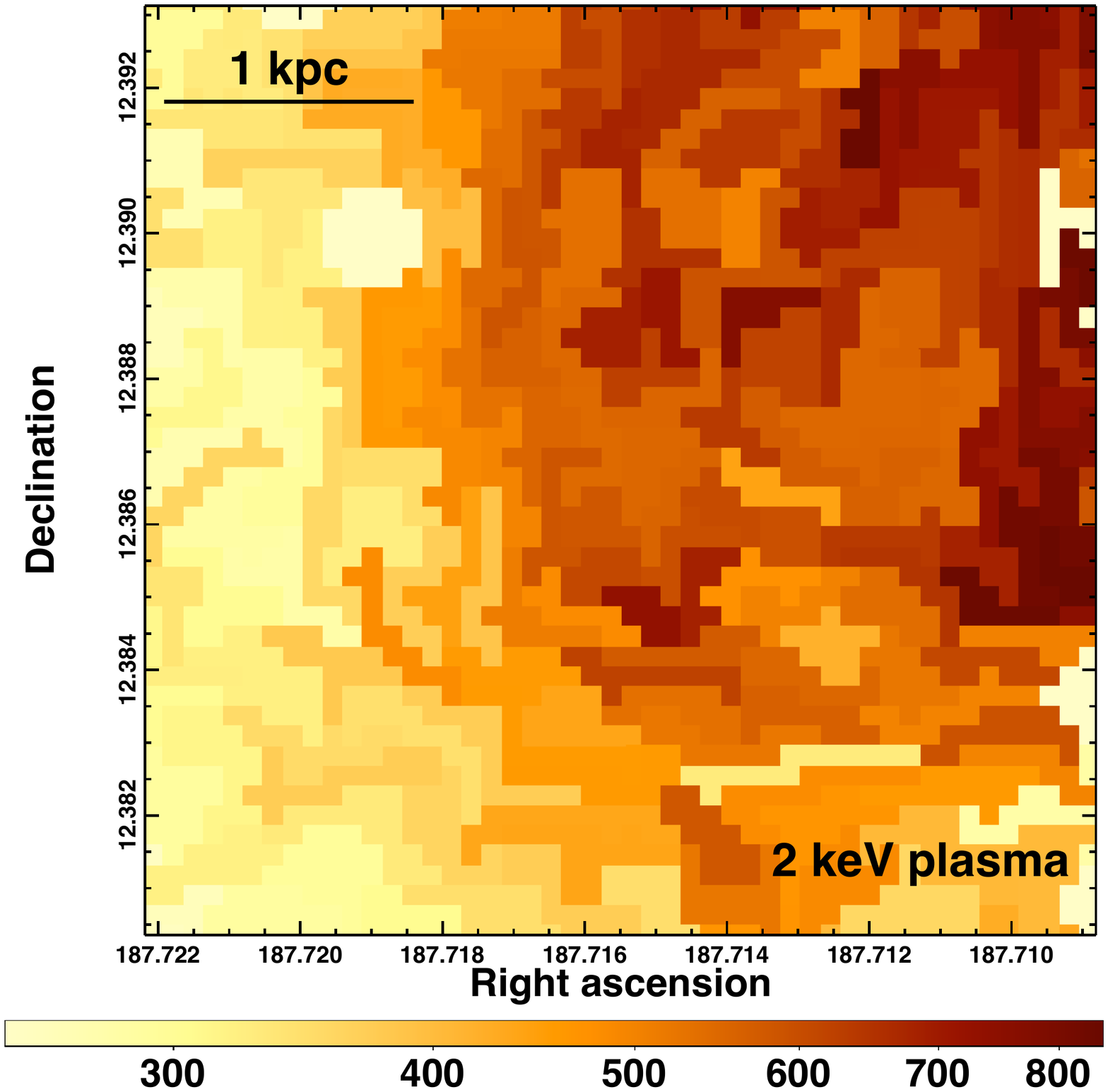}
\end{minipage}
\caption{
FIR [\ion{C}{2}] maps (top panels; we only show spatial bins where the signal-to-noise ratio of the integrated [\ion{C}{2}] flux is greater than 2) and H$\alpha$+[\ion{N}{2}], FUV, and X-ray images of the filamentary multiphase gas southeast of the nucleus of M~87. All images show the same $47\times47$ arcsec region of the sky. The projected 
distance of the bright filaments from the nucleus is 38 arcsec (3~kpc). 
{\it Top left panel: } Map of the integrated [\ion{C}{2}] line flux in units of $10^{-14}$~erg~s$^{-1}$~cm$^{-2}$ per 6\arcsec$\times$6\arcsec spaxel obtained with {\it Herschel} PACS. 
{\it Top central panel: } The velocity distribution of the [\ion{C}{2}] emitting gas, in units of km~s$^{-1}$, relative to the systemic velocity of M~87, $v=1307$~km~s$^{-1}$. 
{\it Top right panel:} Map of the velocity dispersion, $\sigma$, of  the [\ion{C}{2}] emitting gas. 
{\it Middle row left panel:} H$\alpha$+[\ion{N}{2}] image obtained with {\it HST} WFPC2. The contours of these filaments are over-plotted on the {\it Herschel} PACS maps in the top row. On this  {\it HST} WFPC2 image, contours of 6~cm radio 
emission from \citet{hines1989} are over-plotted in red.
{\it Middle row central panel:} FUV image showing \ion{C}{4} line emission obtained with {\it HST} ACS/SBC. 
{\it Middle row right panel:} {\it Chandra} soft X-ray image, extracted in the 0.7--0.9~keV band.
{\it Bottom panels: }Spatial distribution of the emission measure, $Y=\int n_{\mathrm{H}}n_{\mathrm{e}}\mathrm{d}V$, (in $10^{58}$~cm$^{-3}$~arcsec$^{-2}$) of the 0.5~keV {\it (bottom left)}, 1.0~keV {\it (bottom central)}, 2.0~keV 
{\it (bottom right)} plasma detected at 99.7 per cent confidence.
} 
\label{c2}
\end{figure*}

\section{Observations and data analysis}
\label{analysis}
\subsection{Far Infrared spectroscopy with {\it Herschel} PACS}

We observed the FIR cooling lines of [\ion{C}{2}]$\lambda157\mu$m, [\ion{O}{1}]$\lambda63\mu$m, and [\ion{O}{1b}]$\lambda145\mu$m with the PACS integral-field spectrometer \citep{poglitsch2010} on the {\it Herschel} space observatory 
\citep{pilbratt2010}. The observations, with a duration of 10880 seconds,  were performed on January 4th, 2012 (Obs. ID: 1342236278). Table~\ref{obs} gives a summary of the observations. Columns list the observed lines, their rest frame 
wavelengths, observation durations, the observed line fluxes and the $2\sigma$~upper limits, the spectral resolution full-width-at-half-maximum (FWHM) of the instrument at the given wavelength, the observed FWHM, and the observed line 
shift with respect to the systemic velocity of M~87 ($v=1307$~km~s$^{-1}$). 

The observations were taken in line spectroscopy mode with chopping-nodding to remove the telescope background, sky background and dark current. A chopper throw of 6\arcmin\ was used. All three line observations were taken in pointed 
mode centered on ($\alpha$, $\delta$)=(12:30:51.5, +12:23:07) (J2000).

The observations were reduced using the {\tt HIPE} software version 8.2.0, using the PACS {\tt ChopNodLineScan} pipeline script for pointed observations. This script processes the data from level 0 (raw channel data) to level 2 (flux 
calibrated spectral cubes). 

During the final stage of reduction the data were spectrally and spatially rebinned into a 5$\times$5$\times\lambda$ cube. In the following we will refer to this cube as the re-binned cube. Each spatial pixel, termed {\it spaxel}, in this cube has 
a size of 9.4\arcsec$\times$9.4\arcsec. The cube thus provides us with a field of view (FoV) of 47\arcsec$\times$47\arcsec.

For the wavelength re-gridding we set the parameters \textit{oversample} and \textit{upsample} equal to 2 and 1 respectively. This means that one spectral bin corresponds to the native resolution of the PACS instrument (see the PACS Data 
Reduction Guide\footnote{http://herschel.esac.esa.int/twiki/pub/Public/PacsCalibrationWeb/\\PDRG\_Spec\_May12.pdf} for further information). Larger values for both these two parameters were investigated and did not change the results.

For the [\ion{C}{2}] line we projected the re-binned cube onto the sky using the {\tt specProject} task in HIPE and the {\tt hrebin} task in {\tt IDL}. In the following we will refer to this cube as the projected cube. Upon projecting the observed [\ion{C}{2}] data from the telescope frame to the sky we have chosen a resolution of 6\arcsec\ in order to Nyquist sample the beam, the FWHM of which is 12\arcsec\ at the observed wavelength of the line. We only consider spatial bins where the signal-to-noise ratio of the integrated [\ion{C}{2}] flux is greater than 2.

\subsection{{\it HST} H$\alpha$+[\ion{N}{2}] and FUV data}
\label{HSTanalysis}

To study the detailed morphology of the filamentary emission-line nebulae, we have also analyzed H$\alpha$+[\ion{N}{2}] images taken with the Wide Field Planetary Camera 2 (WFPC2) through the F658N filter for 2700 seconds (proposal 
ID: 5122). The central wavelength of the filter is at 6591 \AA\ and its bandwidth is 29 \AA. It transmits the H$\alpha$ line at $\lambda=6563$\AA, with transmission $T= 0.80$, and two [\ion{N}{2}] lines at $\lambda=6584$\AA, with $T=0.20$ and 
at $\lambda=$6548\AA, with $T=0.76$, at the systemic velocity of M~87 ($v=1307$~km~s$^{-1}$).  We subtracted the emission of the underlying stellar population of the galaxy using a red continuum image taken with WFPC2 through the 
filter F814W (proposal ID: 5941). 

A FUV image was obtained using the Advanced Camera for Surveys Solar Blind Channel (ACS/SBC) (proposal ID: 11861) through the filter F150LP (1630 seconds), which covers the \ion{C}{4} line at $\lambda=1549$ \AA\ \citep[see][]
{sparks2009}.

\subsection{{\it Chandra} X-ray data}
Extensive \chandra\ X-ray observations of M~87 were made between July 2002 and November 2005 using the Advanced CCD Imaging Spectrometer (ACIS). The total net exposure time after cleaning is 574~ks. The data reduction is 
described in  \citet{million2010b}. We extracted six background subtracted, flat-fielded, narrow band images between 0.3~keV and 2.0 keV with a $0.492\times0.492$~arcsec$^2$ pixel scale (raw detector pixels).  
To measure the detailed properties of the soft X-ray emission, we also extracted spectra from regions determined using the Contour Binning algorithm \citep{sanders2006b}, as described in \citet{werner2010}. Spectral modeling has been 
performed with the {\tt{SPEX}} package \citep[][]{kaastra1996} in the 0.5--7.0 keV band.

\subsection{Long slit optical spectra}

Long slit spectroscopy was performed using the Intermediate dispersion Spectrograph and Imaging System (ISIS) at the 4.2~m William Herschel Telescope on the island of La Palma on February 10, 2011. The total on-source integration time 
was 3500~s, split into 7 exposures, for the blue and red arm simultaneously.
The data were reduced using standard {\tt IRAF} (Image Reduction and Analysis Facility) procedures. The CCD images were bias subtracted, 
flat fielded, wavelength calibrated, flux calibrated, and merged. 
Spectra were shifted into the heliocentric velocity frame. Cosmic rays were removed with the
{\tt CRNEBULA} subroutine, designed for diffuse objects. Blue and red arm images were treated separately during the whole analysis.

\begin{figure*}
\begin{minipage}{0.48\textwidth}
\includegraphics[width=8cm,clip=t,angle=0.]{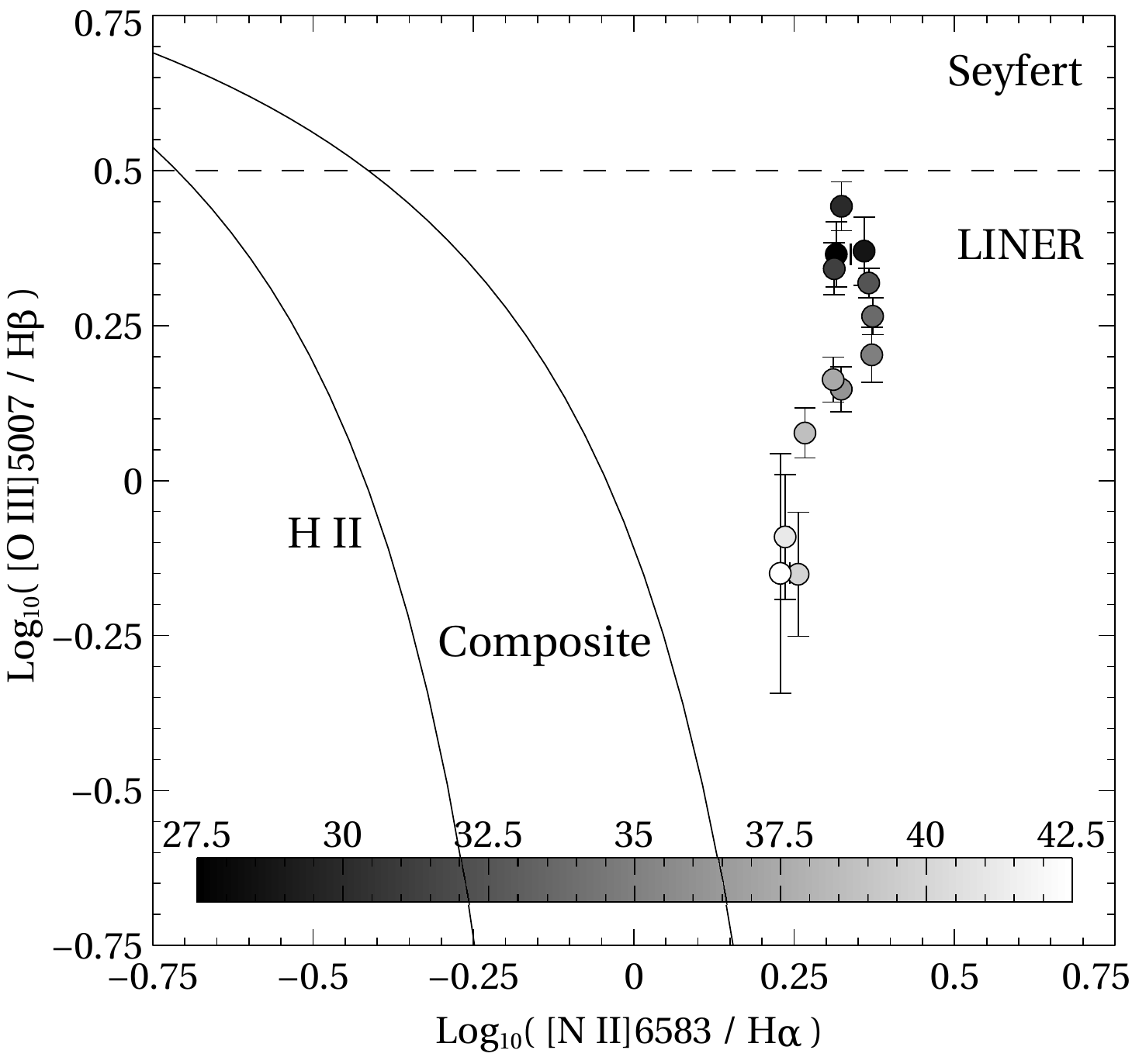}
\end{minipage}
\begin{minipage}{0.48\textwidth}
\includegraphics[width=8cm,clip=t,angle=0.]{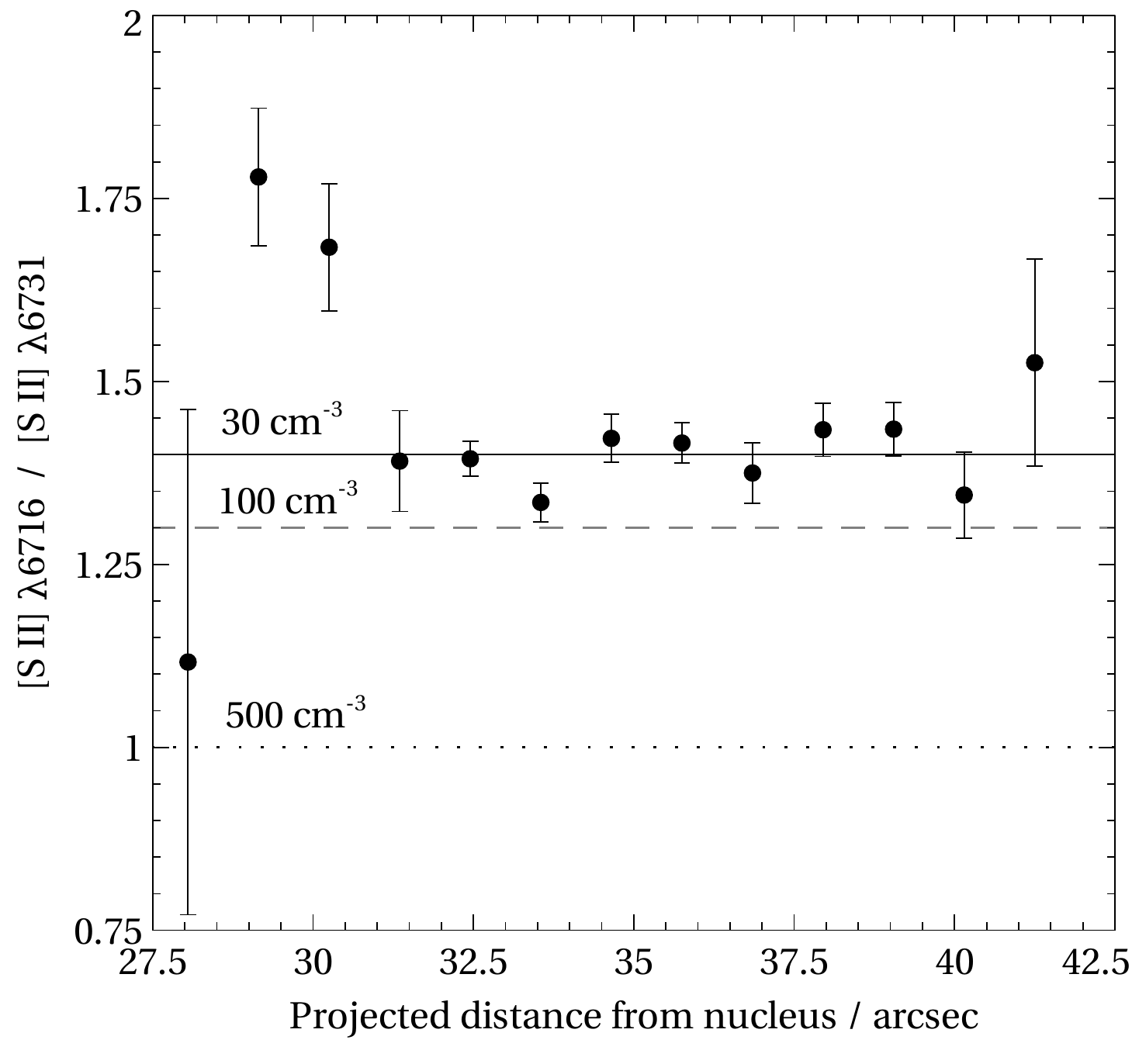}
\end{minipage}
\caption{{\it Left panel:} Optical diagnostic diagram \citep[a BPT diagram,][]{baldwin1981} showing the [\ion{O}{3}]/H$\beta$ against the [\ion{N}{2}]/H$\alpha$ flux observed in the southern filament at different radii from the nucleus. The 
brightening shades of gray indicate increasing projected distance from the nucleus in arcsec. While the side of the filament closest to the nucleus has spectra similar to Seyfert like AGN, at larger radii the gas has a LINER-like emission 
spectrum. The solid and dashed lines are from \citet{kewley2006} and \citet{osterbrock2006}, respectively. {\it Right panel:} The measured [\ion{S}{2}]$\lambda6716$/[\ion{S}{2}]$\lambda6731$ line ratios, a good probe of gas density, as a 
function of radius. The full, dashed, and dotted lines indicate ratios corresponding to electron densities of $n_{\rm e}=30$~cm$^{-3}$, $n_{\rm e}=100$~cm$^{-3}$, and $n_{\rm e}=500$~cm$^{-3}$, respectively \citep{osterbrock2006}. } 
\label{wht}
\end{figure*}

\section{Results}
\label{results}

\subsection{Cold and warm gas phases}

We detect the filamentary emission-line nebulae across all wavebands studied, from the FIR to the soft X-rays. Using {\it Herschel} PACS, we detect [\ion{C}{2}]$\lambda$157$\mu$m line emission (see Fig.~\ref{line}) and determine $2\sigma$ 
upper limits for the [\ion{O}{1}]$\lambda63\mu$m and [\ion{O}{1b}]$\lambda145\mu$m cooling lines. The properties of the FIR lines, spatially integrated over the 5$\times$5 spaxel (47\arcsec$\times$47\arcsec) {\it Herschel} PACS field of 
view, are summarized in Table~\ref{obs}. 
The $2\sigma$ upper limits on the fluxes of the [\ion{O}{1}]$\lambda63\mu$m and [\ion{O}{1b}]$\lambda145\mu$m lines given in Table~\ref{obs} were determined assuming that their velocity widths are equal to the FWHM of the [\ion{C}{2}] 
line.

The [\ion{C}{2}]$\lambda$157$\mu$m line emission is extended and spatially coincident with the H$\alpha$+[\ion{N}{2}] filaments (see the top left panel of Fig.~\ref{c2}). We see two prominent filamentary structures, which we will call the `southern filament' and the `northern filament'. The brightest southern filamentary structure falls on the central spaxel of the PACS spectrometer. The [\ion{C}{2}] flux detected from the central spaxel ($9.4\times9.4$~arcsec$^2$) in the rebinned cube is $(8.5\pm0.7)\times10^{-15}$~erg~s$^{-1}$~cm$^{-2}$.  The top central panel of Fig.~\ref{c2} shows a map of the velocity distribution of the [\ion{C}{2}] line emitting gas with respect to the velocity of M~87. This image shows an interesting dichotomy, with the northern filaments receding at about +140~km~s$^{-1}$ and the bright southern filament (most of the emission of which falls on the central $9.4\times9.4$~arcsec$^2$ spaxel of the detector) moving in the opposite direction along our line of sight at $-123\pm5$~km~s$^{-1}$, with respect to M~87. These velocities show good agreement with the radial velocities of the corresponding optical filaments measured using long-slit spectroscopy \citep{sparks1993}. The top right panel of Fig.~\ref{c2} shows a map of the velocity dispersion of the [\ion{C}{2}] line emitting material. While the northern, receding filaments show a significant velocity dispersion of $\sigma={\rm FWHM}/2.355\sim124$~km~s$^{-1}$ (FWHM$=\sqrt{377^2-240^2}=291$~km~s$^{-1}$, corrected for the instrumental resolution FWHM of 240 km~s$^{-1}$), the velocity dispersion measured in the southern filaments, $\sigma\sim55$~km~s$^{-1}$ (FWHM=$\sqrt{273^2-240^2}=130$~km~s$^{-1}$), is just marginally above the instrumental resolution.

The middle row left panel of Fig.~\ref{c2} shows the H$\alpha$+[\ion{N}{2}] image obtained with {\it HST} WFPC2. Contours of 6~cm radio emission from \citet{hines1989} are over-plotted, showing that in projection the side of the southern 
filament closest to the nucleus overlaps with the radio lobes. The narrowest resolved filaments of H$\alpha$+[\ion{N}{2}] emission have a diameter of only 0.4 arcsec, which corresponds to 32~pc. The total count rate measured in our narrow 
band image from the bright southern filamentary structure within a 15.7\arcsec$\times$5.7\arcsec\ box-like region centered at ($\alpha,\delta$)=(12:30:51.48,+12:23:07.33) is 6.94~counts~s$^{-1}$. Assuming an [\ion{N}{2}]$
\lambda6548$/H$\alpha$ flux ratio of 0.81 and [\ion{N}{2}]$\lambda6584$/H$\alpha$=2.45 \citep{ford1979} and taking into account the filter throughput at the wavelengths of the lines (see Sect.~\ref{HSTanalysis}), the H$\alpha$ flux of the 
southern filament is $f_{\rm H\alpha}=9.0\times10^{-15}$~erg~s$^{-1}$~cm$^{-2}$. 

\citet{sparks2009} reported that the H$\alpha$+[\ion{N}{2}] emission-line nebulae are co-spatial with FUV emission (see the middle row central panel of Fig.~\ref{c2}). Using subsequent spectroscopic measurements with the {\it HST} Cosmic 
Origin Spectrograph, \citet{sparks2012} showed that the FUV emission in the northern filaments is due to \ion{C}{4} line emission. We find that the \ion{C}{4} flux, measured within the same aperture as the H$\alpha$+[\ion{N}{2}] flux above, is 
$f_{\rm CIV}=1.2\times10^{-14}$~erg~s$^{-1}$~cm$^{-2}$. 

The optical spectral properties of the southern filament show an intriguing rapid decrease in the [\ion{O}{3}]/H$\beta$ ratio with increasing distance from the nucleus, dropping monotonically from $\sim$2.5 at a projected distance of 28 arcsec 
to $\sim$0.7 at 43 arcsec (see Fig.~\ref{wht}). The diagram in the left panel of Fig.~\ref{wht}, showing the [\ion{O}{3}]/H$\beta$ over the [\ion{N}{2}]/H$\alpha$ ratios \citep[a BPT diagram,][]{baldwin1981}, indicates that the side of the filament 
closest to the nucleus, which also overlaps with the radio lobes (see middle row left panel of Fig.~\ref{c2}), has spectra similar to Seyfert like active galactic nuclei (AGN). The [\ion{O}{3}]/H$\beta$ line ratios in this part of the filament are larger than in the nucleus of 
M~87, where we measure a ratio of $\sim1.75$. At larger radii, the gas has a LINER\footnote{Low-ionization nuclear emission-line region \citep{heckman1980}}-like emission spectrum, which is more similar to the emission line spectra 
typically seen in extended filaments around brightest cluster galaxies in cluster cores. 

The right hand panel of Fig.~\ref{wht} shows the measured [\ion{S}{2}]$\lambda6716$/[\ion{S}{2}]$\lambda6731$ line ratio, a good probe of gas density, as a function of radius. The full, dashed, and dotted lines indicate ratios corresponding to 
electron densities of $n_{\rm e}=30$~cm$^{-3}$, $n_{\rm e}=100$~cm$^{-3}$, and $n_{\rm e}=500$~cm$^{-3}$, respectively \citep{osterbrock2006}. The value $n_{\rm e}=30$~cm$^{-3}$ is very close to the low density limit, below which the [\ion{S}{2}] line ratios do not provide useful constraints on gas density. Along most of the filament, the line ratios are close to this limit indicating that the electron density of the $10^4$~K phase is $n_{\rm e}\lesssim30$~cm $^{-3}$. At radii between 29--30.5 arcsec, the measured line ratios are outside of the low density limit. The data point closest to the nucleus, at 28 arcsec, indicates a possible increase in density, but with large error bars.

\subsection{The soft X-ray emission}

\citet{young2002} and \citet{sparks2004} found good spatial correlation between H$\alpha$+[\ion{N}{2}] line emission and soft X-ray emission. Here, we have analyzed almost four times as much {\it Chandra} data, which provides excellent 
photon statistics in the regions of interest. 
We extracted six narrow band images between 0.3~keV and 2.0 keV and found that the filaments stand out clearly in the 0.7-0.9 keV band (see the middle row right panel of Fig.~\ref{c2}). 
The spatial correlation between the 0.7--0.9~keV X-rays and the H$\alpha$+[\ion{N}{2}] line emission is remarkable: everywhere, where we see H$\alpha$+[\ion{N}{2}] emitting gas, we also see an excess of soft X-rays. On the other hand, in the 0.5--0.7~keV and 0.9--1.2 keV bands the filaments do not stand out clearly against the surrounding emission, indicating that most of their X-ray emission comes from the \ion{Fe}{17} and \ion{Fe}{18} lines in the 0.7--0.9 keV band. The excess X-ray flux of the filaments in this band is $1.1\times10^{-14}$~ergs~s$^{-1}$~cm$^{-2}$ and their average X-ray surface brightness is $1.2\times10^{-16}$~ergs~s$^{-1}$~cm$^{-2}$~arcsec$^{-2}$.

Assuming that the soft X-ray emission is of a thermal origin, we follow the analysis of  \citet{werner2010} and fit to each spatial bin a model consisting of collisionally ionized equilibrium plasmas at three fixed temperatures (0.5~keV, 1.0~keV, 
and 2.0~keV), with variable normalizations and common metallicity. The bottom panels of Fig.~\ref{c2} show the spatial distributions of the emission measures of the individual temperature components. The 0.5~keV component closely 
follows the distribution of the cold gas phases. This component is required to fit the \ion{Fe}{17} and \ion{Fe}{18} lines in the 0.7--0.9~keV band, which are only present in regions where H$\alpha$+[\ion{N}{2}] emission has also been 
detected. The spatial correlation between H$\alpha$+[\ion{N}{2}]  and the softest X-ray emitting component is, however, not perfect: the ratio of H$\alpha$+[\ion{N}{2}] flux to the $\sim$0.5~keV emission component is $\sim5$ times larger in 
the southern than in the northern filament. The 1~keV component is enhanced as well in the regions where cold gas is present, but it is spatially more extended and is also seen in many regions where cold gas is not detected \citep[see][]
{werner2010}. The spatial distribution of the 2~keV plasma does not correlate with the H$\alpha$+[\ion{N}{2}] emission.

Because the soft X-ray line emission observed in the filaments may be due to cooling gas, we fitted the spectra with a model consisting of a single-temperature plasma in collisional ionization equilibrium plus an isobaric cooling flow, which 
models cooling between two temperatures at a certain metallicity and mass deposition rate. The upper temperature and the metallicity of the cooling flow component were tied to the temperature and metal abundance of the single-temperature plasma. Both components were absorbed by a Galactic absorption column density $N_{\rm H}=1.9\times10^{20}$~cm$^{-2}$ \citep{kalberla2005}. Fixing the lower temperature cut-off of the cooling flow model to $kT_{\rm low}=0.5$~keV, we obtained a best fit upper temperature of $kT_{\rm up}=1.90\pm0.04$~keV, a mass deposition rate of $\dot{M}=(2.46\pm0.13)\times10^{-2}~M_{\odot}$~yr$^{-1}$, and a metallicity of $Z=1.46\pm0.09$~Solar \citep[relative to the 
Solar values by ][]{grevesse1998}. Even though the spectra indicate that above $kT\sim0.5$~keV the gas may be cooling, the spectral signatures of gas cooling below 0.5~keV - soft X-ray emission, including the \ion{O}{7} line - are missing 
\citep[see also the high resolution spectra in][]{werner2006b,werner2010}.

In order to test whether the missing soft X-ray flux might have been absorbed by the cold gas in the filaments, we fixed the mass deposition rate in the model to the best fit value obtained with a lower temperature cut-off set to 0.5~keV, 
extended this cut-off down to a 1000 times lower temperature of 0.5~eV (the lowest value allowed by the model). Furthermore, we assumed that the cooling X-ray plasma is intermixed with the cold gas and the cooling happens in 13 different regions along our line-of-sight separated from each other by intrinsic cold absorbers (the emission from the first cooling region is affected by one absorber, from the second by two absorbers, from the third by three etc.). The different cooling regions are assumed to have the same cooling rate and the different absorbers are assumed to have the same hydrogen column density. Fitting this model to the data, we find an integrated intrinsic hydrogen column density of $N_{\rm H}=(1.58\pm0.21)\times10^{21}$~cm$^{-2}$ and an absorbed bolometric flux of $3.9\times10^{-14}$~erg~s$^{-1}$~cm$^{-2}$. 

\section{Discussion}
\label{discussion}

\subsection{Magnetized filaments of multi-phase material}

In the filamentary emission-line nebulae to the southeast of the nucleus of M~87, we detect gas spanning a temperature range of at least 5 orders of magnitude, from $\sim100$~K to $\sim10^7$~K. 
The [\ion{C}{2}], H$\alpha$+[\ion{N}{2}], and \ion{C}{4} emitting gas phases are consistent with being co-spatial, forming a multi-phase medium. Soft X-ray emission in the 0.7--0.9~keV band is also always present in regions where H$\alpha$+
[\ion{N}{2}] emission is seen. Its spectral shape is consistent with line emission of $\sim$0.5~keV thermal plasma, previously detected in the {\it XMM-Newton} high resolution reflection grating spectra integrated over the central region of M~87 
\citep{werner2006b,werner2010}. We see no evidence of X-ray emitting gas with temperature $kT<0.5$~keV.  

The [\ion{C}{2}] 158 $\mu$m line is about 1500 times stronger than the CO (1$\rightarrow$0) rotational line in normal galaxies and Galactic molecular clouds, and 6300 times more intense in starburst galaxies and Galactic star forming 
regions \citep{crawford1985,stacey1991}. 
Given a [\ion{C}{2}] flux of $8.5\times10^{-15}$ erg~s$^{-1}$~cm$^{-2}$ observed from the brightest region of the southern filament, the expected CO (1$\rightarrow$0) flux is 1.3--$5.3\times10^{-18}$~erg~s$^{-1}$~cm$^{-2}$, consistent 
with the upper limit of $2\times10^{-17}$~erg~s$^{-1}$~cm$^{-2}$ for CO (1$\rightarrow$0) at 2.6~mm observed in the same region \citep{salome2008}. Using the standard CO luminosity to H$_2$ conversion factors the inferred range of CO 
luminosities corresponds to an H$_2$ mass of 0.4--$2\times10^6~M_\odot$. However, because the [\ion{C}{2}]/CO line ratios in these filaments may be different from those in star-forming galaxies and their heating mechanism may be different from that in more normal molecular clouds, their true molecular mass may be outside of this range.

Assuming that all the observed gas phases are in thermal pressure equilibrium with the ambient ICM \citep[assuming $p=0.11-0.22$~keV cm$^{-3}$, for a distance range of $r=3-6$~kpc from the nucleus depending on the position of the filament along our line of sight,][]{churazov2008}, the $\sim$ 100~K [\ion{C}{2}] emitting gas forms a network of narrow filaments with densities of the order of $(1.3-2.6)\times10^4$~cm$^{-3}$ (higher than the critical density of $3\times10^3$~cm$^{-3}$ above which the gas is in local thermal equilibrium with the level populations determined by collisions) with volume filling fraction of $f_{\rm V}\sim(1-2)\times10^{-4}$ in the bright southern filament. Although the presence of \ion{O}{1} 6300~\AA\ lines in the optical spectra \citep[see also][]{ford1979} indicates the existence of large partially ionized regions in the filaments, the presence of [\ion{S}{2}] lines most probably indicates a fully ionized $\sim10^4$~K phase surrounding the colder gas. The density of this H$\alpha$+[\ion{N}{2}] emitting gas, assuming thermal pressure equilibrium with the ambient ICM, should be $\sim(1.3-2.6)\times10^2$~cm$^{-3}$.

The  [\ion{S}{2}]$\lambda6716$/[\ion{S}{2}]$\lambda6731$ line ratios, however, indicate much lower particle densities in the $10^4$~K phase ($n\sim2n_{\rm e}\lesssim60$~cm$^{-3}$) implying the presence of additional significant non-thermal pressure components in the filaments, such as turbulence and magnetic fields. Assuming isotropic micro-turbulence with a characteristic velocity $v_{\rm turb}$ in a medium with a sound speed $c_{\rm s}$, the turbulent pressure support will be $p_{\rm turb}=1/2\, \gamma\, p_{\rm therm}\, M^2$, where $\gamma$ is the adiabatic index assumed to be 5/3 and $M=v_{\rm turb}/c_{\rm s}$ is the Mach number. However, assuming characteristic turbulent velocities of the order of the sound speed in the $10^4$~K phase ($c_{\rm s}=15$~km~s$^{-1}$), the sum of thermal and turbulent pressure,  $p_{\rm turb}+p_{\rm therm}=0.09$~keV~cm$^{-3}$ ($1.4\times10^{-10}$~dyne~cm$^{-2}$), is still smaller than the surrounding ICM pressure. If the turbulence in the H$\alpha$ emitting phase is subsonic, then the additional pressure will be provided by magnetic fields. A magnetic pressure of $p_{\rm mag}=B^2/8\pi\sim(0.3-2.1)\times10^{-10}$~dyne~cm$^{-2}$, needed to keep the filaments in pressure equilibrium with the surrounding ICM, requires magnetic fields of $B=28-73~\mu$G. This is similar to the values inferred using arguments based on the integrity of H$\alpha$+[\ion{N}{2}] filaments in the Perseus Cluster \citep{fabian2008}. Radio observations of the Faraday rotation measure in a number of cooling cores revealed magnetic field strengths of 10--25~$\mu$G \citep{taylor2001,taylor2007,allen2001,feretti1999}.  Significant imbalance in thermal pressure of the gaseous filaments in cool-core clusters has previously also been inferred between the warm molecular hydrogen gas seen in NIR  ($nT\sim10^{8-9}$~cm
$^{-3}$~K) and the ionized gas phases seen in the optical ($nT\sim10^5$~cm$^{-3}$~K), emphasizing the need for dynamic models for the filaments \citep{jaffe2001,oonk2010}.

The smallest resolved diameter of the H$\alpha$+[\ion{N}{2}] filaments is 32~pc, about half of the 70~pc diameter observed in the more distant Perseus Cluster \citep{fabian2008}. However, even the narrowest resolved filaments are likely to 
consist of many smaller strands. They are surrounded by hotter  $10^5$~K \ion{C}{4} emitting gas. The space between these filaments is filled by X-ray emitting gas. If the soft X-ray line emission is due to  $\sim$0.5~keV gas, then based on 
our three-temperature fits (see Sect.~\ref{results}) and assuming thermal pressure equilibrium with the ambient ICM, its mass in the southern filament is $\sim10^5~M_{\odot}$ and its emitting volume is $\sim$0.016~kpc$^3$. 

\subsection{The energy source of the filaments}

\citet{sparks2009} show that no main sequence stars of early spectral type are present in the filaments and therefore neither the FUV nor the H$\alpha$+[\ion{N}{2}] emission can be due to photoionization by hot UV emitting stars. 
The [\ion{O}{3}]/H$\beta$ emission line ratios are high on the side of the southern filament closest to the nucleus, which also overlaps with the radio lobes (see middle row left panel of Fig.~\ref{c2}), and then decrease rapidly with 
increasing distance from the nucleus. The fact that the line ratios in the filament reach values which are higher than those measured in the nucleus argues against photoionization by the AGN. It seems more likely that the increased 
[\ion{O}{3}]/H$\beta$ ratios are due to internal shocks produced as the radio lobes push against the cold gas. These shocks seem to affect only a part of the filament and the emitted H$\alpha$ flux does not decrease with the decreasing [\ion{O}{3}]/H$\beta$ emission line ratios. Therefore, they can be ruled out as energy source powering the emission of these nebulae.

 \citet{ferland2009} showed that the broad band (optical to mm) emission-line spectra of nebulae around central galaxies of cooling core clusters most likely originate in gas exposed to ionizing particles. Such ionizing particles can be either 
relativistic cosmic rays or hot particles penetrating into the cold gas from the surrounding ICM.
Saturated conduction from the hot into the cold phase of the southern filament would produce a heat flux of $5\phi p c_{\rm{s}} \sim 0.1$~erg~s$^{-1}$~cm$^{-2}$, where $\phi\sim1$ accounts for the uncertain physics, $p=0.22$~keV~cm
$^{-3}$ is the gas pressure and $c_{\rm s}=510$~km~s$^{-1}$ is the sound speed in the hot 1~keV phase \citep{cowie1977,fabian2011}. Following the argument of \citet{fabian2011}, given a typical H$\alpha$ surface brightness of $
\sim2\times10^{-16}$~erg~s$^{-1}$~cm$^{-2}$~arcsec$^{-2}$, corresponding to an emitted flux of  $1.1\times10^{-4}$~erg~s$^{-1}$~cm$^{-2}$ and assuming that the total broad band emitted flux is 20 times larger $2.2\times10^{-3}$~erg~s
$^{-1}$~cm$^{-2}$, the required efficiency with which impinging thermal particles penetrate and excite the cold gas is of the order of $f=0.022$ ($f=0.044$ for a filament at $r=6$~kpc from the nucleus), which is reasonable. However, these 
particles must somehow overcome the obstacle presented by the magnetic fields believed to be at the interface of the cold and hot gas phases. 

\citet{werner2010} show that all of the bright H$\alpha$+[\ion{N}{2}] and UV filaments are seen in the downstream region of a $<3$~Myr old AGN induced $M\gtrsim1.2$ shock front \citep{million2010b} that seems to have just passed the 
filaments. Based on these observations they propose that shocks induce shearing around the cold, filamentary, dense gas, thereby promoting mixing of the cold gas with the ambient hot ICM via instabilities \citep[e.g.][]{friedman2012}. 
Instabilities and turbulence may also be driven by shearing motions between the hotter X-ray gas and the filaments, as they move through the ambient ICM, both as they are being uplifted and as they fall back. This type of more continuous 
shear induced mixing could perhaps be in the long term more important for energizing the filaments than shocks.  \citet{fabian2011} propose that the sub-mm through UV emission seen in NGC~1275 in the core of the Perseus Cluster is due 
to the hot ICM efficiently penetrating the cold gas through reconnection diffusion \citep{lazarian2010,lazarian2011}. The same process may also be operating in the filaments of M~87 with the fast reconnection induced by shearing instabilities 
and turbulence. Shearing and magnetic reconnection are, however, not essential to bring the cold gas into contact with the hot ICM. The support from magnetic fields tied to the surrounding ionized phase may become unstable if the layer of dense cold neutral gas is deep enough, and therefore it may fall through into direct contact with the hot gas. The hot ICM particles penetrating the cold gas then produce secondary electrons that excite the observed FIR through UV emission as shown by \citet{ferland2009}. Mixing of the cold and hot gas also naturally account for the 
presence of the $\sim10^5$~K intermediate-temperature gas phase (geometric mean of the hot and cold gas temperatures) predicted by the mixing layers calculated by \citet{begelman1990}, although mixing layers have most probably a 
more complicated temperature structure \citep{esquivel2006}. As the X-ray emitting gas cools through mixing, the mass of the southern filament may grow by as much as $4\times10^{-2}~M_{\odot}$~yr$^{-1}$.

The particle heating model of \citet{ferland2009}, calculated considering optically thin emission from a cell of gas, predicts [\ion{O}{1}]/[\ion{C}{2}]$~\sim21$ and is inconsistent with the $2\sigma$ upper limit of 7.2 observed in M~87. 
The ratios observed in the Perseus and Centaurus clusters, and in Abell~1068 and 2597 are even lower. This discrepancy has been discussed in a recent study of NGC~1275 by \citet{mittal2012}, who conclude that the line ratio suggests 
that the lines are optically thick, implying a large reservoir of cold atomic gas, which was not accounted for in previous inventories of the filament mass. The same can be concluded for M~87, where the large optical depth of the FIR lines 
(Canning et al. in prep.) and the intrinsic absorption inferred from the X-ray data (see below) imply significant reservoirs of cold atomic and molecular gas. 

\subsection{X-ray line emission from cold gas due to charge exchange and inner shell ionization?}
\label{cxc}

As ions from the hot ICM penetrate into the filaments, they pick up one or more electrons from the cold gas. The impinging ions thus get into excited states and their deexcitation may produce X-ray radiation. Astrophysical X-ray emission due 
to this `charge exchange' process was first discovered in the comet Hyakutake \citep{lisse1996,cravens1997} and since then it has been identified as the mechanism responsible for  X-rays from planets \citep[e.g.][]
{gladstone2002,dennerl2002}, the geocorona \citep{wargelin2004,fujimoto2007}, the heliosphere \citep{robertson2001}, and interstellar medium in particular at interfaces between hot gas and cool clouds \citep[see the review by][and 
references therein]{dennerl2010}. Recently, charge exchange has also been found to possibly contribute significantly to the soft X-ray line emission from star-forming galaxies \citep{liu2011,liu2012}. \citet{fabian2011} conclude that in the 
Perseus Cluster charge exchange may account for a few per cent of the soft X-ray emission from the filaments. They also speculate that if the filaments are composed of many small strands, the same ion might pass from the hot into the cold 
and back to the hot phase repeatedly, getting collisionally ionized while passing through the hot ICM, thus enhancing the charge exchange emission to the observed level. 

After hydrogen and helium, the most abundant ions penetrating the cold gas are O$^{8+}$, C$^{6+}$, N$^{7+}$, Ne$^{10+}$, Mg$^{12+}$, Mg$^{11+}$ (with concentrations $8.4\times10^{-4}$, $3.6\times10^{-4}$, $1.1\times10^{-4}$, 
$1.1\times10^{-4}$, $2.5\times10^{-5}$, $1.07\times10^{-5}$ relative to hydrogen by number). Charge transfer onto these ions is expected to produce mainly hydrogen and helium like line emission in the X-ray band. In the 1~keV plasma 
surrounding the filaments, 23\% of Fe is in the form of Fe$^{20+}$, another 23\% in Fe$^{21+}$, 18\% in Fe$^{19+}$, and 17\% in Fe$^{22+}$, with relative concentrations smaller than $1.1\times10^{-5}$. The charge exchange line fluxes 
produced by Fe will therefore be significantly lower, with most of the line flux in \ion{Fe}{19} to \ion{Fe}{22} lines, in the spectral band above 0.9~keV. Moreover, laboratory measurements of X-ray spectral signatures of charge exchange in L-
shell Fe ions by \citet{beiersdorfer2008} found that the $n\ge4 \rightarrow n=2$ transitions are strongly enhanced relative to $n=3 \rightarrow n=2$, shifting the line emission to higher energies. Because charge transfer processes and the 
corresponding X-ray emission are expected to occur inside of the dense cold clouds, absorption may significantly decrease the observed fluxes of the lowest energy X-ray lines such as \ion{C}{6}, \ion{N}{7}, \ion{O}{7}, and \ion{O}{8}. 
The \ion{Ne}{9},  \ion{Ne}{10},  \ion{Mg}{11}, and \ion{Mg}{12} lines in the 0.9--1.9~keV band are, however, not expected to be affected by absorption strongly and their enhancement might indicate charge exchange. Given that none of these 
lines are enhanced in the filaments, and no strong charge exchange emission is expected in the 0.7--0.9~keV band, we conclude that the observed soft X-ray emission is not due to charge transfer onto the impinging ions. 

The ions and electrons penetrating from the surrounding hot ICM into the cold filaments will also ionize the inner electron shells of the atoms and molecules in the cold clouds. This inner shell ionization may in principle also contribute to the 
X-ray emission of the filaments, but given the energetics, it is not expected to be significantly stronger than the X-ray emission due to charge exchange.  We would primarily expect to see X-ray lines from few times ionized C, N, O, Ne, and Mg, 
with the Ne and Mg lines in the band above 0.6 keV. Detailed predictions of this process are beyond the scope of this paper. 

Given that the spectral shape of the filaments is consistent with line emission of $\sim$0.5~keV thermal plasma previously resolved within the  central region of M~87 with {\it XMM-Newton} RGS \citep{werner2006b,werner2010} and that 
charge exchange as a significant source of X-ray emission from the filaments can be ruled out, we conclude that the observed soft X-ray emission is most likely due to thermal plasma. We note, however, that although charge exchange is not responsible for the observed excess X-ray emission in the $0.7-0.9$~keV band, the level which is required from the particle heating model is observationally acceptable.

\subsection{AGN uplift induced cooling instabilities or mixing?}

The observed velocity distribution of the [\ion{C}{2}]  (see top central panel of Fig.~\ref{c2}) and optical emission lines \citep{sparks1993} indicates that the northern filaments are currently being uplifted by the counter-jet \citep[which is oriented 
at $<19^{\circ}$ from our line-of-sight,][]{biretta1999} moving away from M~87 with a line-of-sight velocity of $v_{\rm LOS}=+140$~km~s$^{-1}$. The southern filaments, in contrast, are falling back towards M~87 with $v_{\rm LOS}
=-123$~km~s$^{-1}$. 
Assuming uplift at the Keplerian velocity of $\sim400$~km~s$^{-1}$, the southern filament would take $8\times10^6$~yr to get to its current position. Given that this filament is most likely already falling back, its age is $\gtrsim10^7$~yr. 
Uplift of low entropy, metal enriched X-ray emitting gas by buoyantly rising relativistic plasma from the AGN jets of M~87 has been previously inferred based on radio and X-ray data 
\citep{churazov2001,forman2005,simionescu2008a,werner2010}. Optical emission line nebulae also appear to have been drawn out by rising radio bubbles in other nearby systems \citep[e.g. the Perseus and Centaurus clusters,][]
{hatch2006,canning2011}.

The initially uplifted material might have already been a multi-phase mixture of low entropy X-ray emitting plasma and dusty cold gas, as is observed in the core of M~87 \citep{sparks1993}. As the relatively dense, low entropy X-ray emitting 
plasma was removed from the direct vicinity of the AGN jets it might have cooled radiatively, further contributing to the cold gas mass of the uplifted material \citep[see][]{revaz2008}. The multi-phase X-ray emitting gas, with the 0.5~keV phase 
co-spatial with the H$\alpha$+[\ion{N}{2}] filaments and surrounded by a spatially more extended 1~keV plasma, indicates that such cooling may be taking place. Cooling instabilities are predicted to take place in cooling core clusters within 
radii where the ratio of the cooling time to free fall time is $t_{\rm cool}/t_{\rm ff}\lesssim10$ \citep[][this condition is fulfilled at the radii where the M~87 filaments are seen]{sharma2012,mccourt2012,gaspari2012}. AGN induced uplift of 
multiphase gas may facilitate the onset of such local cooling instabilities. Some H$\alpha$+[\ion{N}{2}] and soft X-ray filaments lie at the edges of cavities filled by radio emitting plasma, indicating that they might be sheets of gas seen in 
projection that have been swept up and compressed by the expanding and rising cavities. Even though such compression would first heat the gas, eventually it will accelerate its cooling. 
This picture of radiatively cooling X-ray gas is, however, complicated by the observed temperature floor at 0.5~keV. Several explanations were proposed to address the missing soft X-ray emission, including mixing \citep{fabian2002}, 
anomalous O/Fe abundance ratios, and absorption \citep{sanders2011}.

The mass deposition rate, inferred from the temperature distribution of the X-ray gas in the southern filament above $kT=0.5$~keV, assuming an isobaric cooling flow model, is  $\dot{M}=(2.46\pm0.13)\times10^{-2}~M_{\odot}$~yr$^{-1}$. 
This cooling rate could only be reconciled with the apparent lack of $kT<0.5$~keV plasma if the cold gas in the filaments absorbs the soft X-rays, modifying the observed spectral shape in a way that results in an apparent temperature cut-off. 
The X-ray spectral lines providing the most diagnostic power at these temperatures are the \ion{Fe}{17} and the \ion{O}{7} lines. The ionic fraction of \ion{Fe}{17} is above 0.3 across a relatively large temperature range of 0.2--0.6 keV, but the 
\ion{O}{7} line becomes observable only at $kT<0.4$~keV. Absorption is expected to strongly suppress the \ion{O}{7} line emitted by the coldest gas phases. The \ion{Fe}{17} lines will be affected by absorption to a much lesser extent. 
Therefore when modeling a spectrum that is affected by intrinsic absorption, we may conclude that the cooling stopped at the temperature where we expect to start to see the \ion{O}{7} line\footnote{We note that only a small fraction of the 
observed X-ray emission originates in the filaments and is affected by intrinsic absorption. The dominant emission component from the ambient ICM remains unaffected by cold gas in the filaments, and therefore fitting the spectra with a 
Galactic absorption as a free parameter will not result in a significantly higher best fit $N_{\rm H}$.}. In Sect.~\ref{results}, we show that the required integrated hydrogen column density, distributed in multiple layers (the cooling takes place between the threads) along our line-of-sight  is $N_{\rm H}=(1.58\pm0.21)\times10^{21}$~cm$^{-2}$. If the hot ICM indeed cools radiatively between the cold filaments, then the large area covering fraction of the many small strands may provide a sufficient absorption column to the cooling X-ray emitting 
plasma to explain the apparent temperature floor at $\sim$0.5~keV. 

Soft X-ray photons from the cooling X-ray plasma absorbed by the cold gas would produce energetic photoelectrons which would contribute to the heating and ionization of the cold gas \citep[in a process similar to the secondary electron 
production by cosmic rays or impinging hot ICM particles,][]{ferland2009}. Photoionization of emission line nebulae by X-rays has been considered before \citep[e.g.][]{heckman1989,donahue1991,crawford1992,jaffe1997} and has been 
found insufficient to ionize and power the observed line emission. However, when considering photoionization by primarily soft X-rays, from cooling plasma which surrounds the cold filamentary gas with a small volume filling fraction, the 
efficiency of this process increases significantly \citep{oonkthesis}. 
According to our spectral modeling, where we assumed that all the soft X-ray emission is due to radiative isobaric cooling, intrinsic absorption reduces the observed flux  by $\sim3.9\times10^{-14}$~erg~s$^{-1}$~cm$^{-2}$. If the soft X-ray 
emitting plasma fills the space between the cold filaments and the cold gas absorbs approximately the same amount of energy in all directions, then the total X-ray energy absorbed by the cold gas is $\sim1.3\times10^{39}$ erg~s$^{-1}$. 
With a total emitted energy from UV to sub-mm about 20 times larger than the observed H$\alpha$ luminosity of $3.0\times10^{38}$~erg~s$^{-1}$, X-ray photoionization may, in principle, contribute a significant fraction of the energy needed 
to power the filaments.

The cooling rates above were deduced under the assumption that all the soft X-ray line emitting gas originates from radiative cooling. The $kT\sim0.5$~keV gas could, in principle, also be produced by mixing of the hot ICM with the cold 
filaments. 
More developed mixing might be responsible for the higher ratio of 0.5~keV thermal to H$\alpha$+[\ion{N}{2}] line emission in the southern filament (which is most likely falling back towards M~87), with respect to the northern filament (which is likely younger, currently being uplifted).
Such mixing is, however, not expected to result in a gas with $kT\sim0.5$~keV - it would result in a mixing layer with a wide range of temperatures \citep{esquivel2006}. Therefore, both in the cooling and in the mixing scenario for the origin of 
the soft X-ray emission from the filaments, the explanation of the 0.5~keV cutoff requires intrinsic absorption by cold gas.  

The internal reddening in the optical band calculated from the observed line ratio of H$\alpha$/H$\beta=4$ in the filament \citep{ford1979} is, $E(B-V)=0.28$. Using the Galactic transformation between reddening and column density of 
neutral hydrogen, $N_{\rm H}/E(B-V)=5.8\times10^{21}$~cm$^{-2}$~mag$^{-1}$ \citep{bohlin1978}, we infer an intrinsic neutral hydrogen column density of $N_{\rm H}=1.6\times10^{21}$~cm$^{-2}$ - sufficient to absorb the soft X-ray 
emission from gas with $kT<0.5$~keV. We note that substantial reddening has been observed in a number of H$\alpha$ nebulae in brightest cluster galaxies of cooling core clusters \citep[e.g.][]{allen1995,crawford1999}.

A similar temperature floor at $\sim$0.5~keV is observed in the Perseus Cluster, Centaurus Cluster, 2A~0335+096, Abell~262, Abell~3581, HCG~62, and Abell~2052  
\citep{sanders2007,sanders2008,sanders2009,sanders2009b,deplaa2010} and in the cooling uplifted multi-phase gas in the core of the galaxy cluster S\'ersic~159-03 \citep{werner2011}. The coolest $\sim$0.5~keV gas phases are co-spatial with filamentary optical emission-line nebulae in all of these systems. In the Perseus and Centaurus clusters, copious amounts of FIR line emission has also been discovered, indicating the presence of large reservoirs of cold atomic 
gas \citep{mittal2011,mittal2012}. Following the arguments outlined above, the low temperature X-ray cutoffs observed in those systems may therefore also, at least partly, result from intrinsic absorption.

\section{Conclusions}
\label{conclusions}

We performed a multi-wavelength study of the emission-line nebulae located southeast of the nucleus of M~87. Our main conclusions may be summarized as follows:

\begin{itemize}
\item We detect FIR [\ion{C}{2}] line emission at 158~$\mu$m with {\it Herschel} PACS. The line emission is extended and co-spatial with optical H$\alpha$+[\ion{N}{2}], FUV \ion{C}{4} lines, and soft X-ray emission. 

\item The filamentary nebulae contain multi-phase material spanning a temperature range of at least 5 orders of magnitude, from $\sim100$~K to $\sim10^7$~K. This material has most likely been uplifted by the AGN from the center of M~87. 

\item The thermal pressure of the $10^4$~K phase appears to be significantly lower than that of the surrounding hot ICM indicating the presence of additional turbulent and magnetic pressure in the filaments. If the turbulence in the 
filaments is subsonic then the magnetic field strength required to balance the pressure of the surrounding ICM is $B\sim30-70~\mu$G.

\item The spectral properties of the soft X-ray emission from the filaments indicate that it is due to thermal plasma with $kT\sim0.5$--1~keV, which is cooling by mixing with the cold gas and/or radiatively. Charge exchange can be ruled out as 
a significant source of soft X-rays. 
  
\item Both cooling and mixing scenarios predict gas with a range of temperatures. This is at first glance inconsistent with the apparent lack of X-ray emitting gas with $kT<0.5$. 
However, we show that the missing very soft X-ray emission could be absorbed by the cold gas in the filaments with an integrated hydrogen column density of $N_{\rm H} \sim 1.6\times10^{21}$~cm$^{-2}$, providing a natural explanation for the 
apparent temperature floor to the X-ray emission at $kT\sim0.5$~keV. The internal reddening observed in the optical band indicates a similar level of intrinsic absorption. 

\item The FIR through UV line emission is most likely primarily powered by the ICM particles penetrating the cold gas following a shearing induced mixing process. An additional source of energy may, in principle, be provided 
by X-ray photoionization from cooling X-ray emitting plasma.

\item The relatively small line ratio of [\ion{O}{1}]/[\ion{C}{2}]$<7.2$ indicates a large optical depth in the FIR lines. The large optical depth at FIR and the intrinsic absorption inferred from the X-ray and optical data all imply significant reservoirs 
of cold atomic and molecular gas distributed in filaments with small volume filling fraction, but large area covering factor. 

\end{itemize}

\acknowledgments
This work is based in part on observations made with {\it Herschel}, a European Space Agency Cornerstone Mission with significant participation by NASA. Support for this work was provided by NASA through award number 1428053 issued 
by JPL/Caltech. 
Support for this work was provided by NASA through Einstein Postdoctoral grant numbers PF9-00070 (AS) and PF2-130104 (RvW) awarded by the Chandra X-ray Center, which is operated by the Smithsonian Astrophysical Observatory for 
NASA under
contract NAS8-03060. 
SWA acknowledges support from the U.S. Department of Energy under contract number DE-AC02-76SF00515. MR acknowledges NSF grant AST 1008454.
\bibliography{clusters}

\bibliographystyle{apj}

\end{document}